\documentclass{aa}
\usepackage{graphicx}

\def\ltsim{\raise 2pt \hbox {$<$} \kern-1.1em \lower 4pt \hbox {$\sim$}}
\def\ltapprox{\raise 2pt \hbox {$<$} \kern-1.1em \lower 5pt \hbox {$\approx
$}}
\def\gtsim{\raise 2pt \hbox {$>$} \kern-1.1em \lower 4pt \hbox {$\sim$}}
\def\gtapprox{\raise 2pt \hbox {$>$} \kern-1.1em \lower 5pt \hbox {$\approx
$}}

\begin{document}
\title{Radio emission at the centre of the galaxy cluster Abell 3560: 
evidence for core sloshing?}

\author{T.~Venturi\inst{1}, M.~Rossetti\inst{2,3}, S.~Bardelli\inst{4},
S.~Giacintucci\inst{5,6}, D.~Dallacasa\inst{7}, M.~Cornacchia\inst{7},
N.G.~Kantharia\inst{8}}
\institute
{
INAF -- Istituto di Radioastronomia, via Gobetti 101, I--40129, Bologna, Italy 
\and
Dipartimento di Fisica, Universit\`a degli Studi di Milano, Via Celoria 16,
I--20133, Milano, Italy
\and
INAF -- IASF, Via Bassini 15, I--20133, Milano, Italy
\and
INAF -- Osservatorio Astronomico di Bologna, Via Ranzani 1, I--40126 Bologna, 
Italy
\and
Department of Astronomy, University of Maryland, College Park, MD 20742-2421, 
USA
\and
Joint Space-Science Institute, University of Maryland, College Park,
MD, 20742-2421, USA
\and
Dipartimento di Astronomia, Universit\`a di Bologna, via Ranzani 1, 
I--40126 Bologna, Italy
\and
National Centre for Radio Astronomy, Tata Institute of Fundamental Research,
Post Bag 3, Ganeshkind, Pune 411007, India
}

\date{Received 00 - 00 - 0000; accepted 00 - 00 - 0000}

\titlerunning{Core sloshing at the centre of Abell~3560?}
\authorrunning{Venturi et al.}

\abstract
{The present paper deals with the interplay between the radio emission 
associated with the dominant galaxy in clusters and the properties of the 
surrounding intracluster medium (ICM), studied on the basis of its
X--ray emission.} 
{Previous radio observations of the galaxy cluster A\,3560, located in the
Shapley Concentration core, revealed the presence of complex radio emission 
associated with the brightest cluster member.
To understand the origin of such radio emission we performed a detailed 
multiwavelength study, with high quality proprietary data in the radio and 
X--ray bands, and by means of optical data available in the literature.}
{We observed the cluster with the Giant Metrewave Radio Telescope, the Very 
Large Array and the Australia Telescope Compact array at 240 and 610 MHz, 1.28,
1.4, 2.3, 4.8 and 8.4 GHz, and performed a detailed morphological and 
spectral study of the radio emission associated with the brightest cluster
galaxy (BCG). Furthermore, 
we observed the cluster with the $XMM$--$Newton$ and $Chandra$ 
observatories, to derive the properties of the intracluster gas.
Finally, we made use of literature data to obtain the bidimensional 
distribution of the galaxies in the cluster.}
{The radio emission, associated with the north--eastern nucleus of the 
dumb--bell BCG, is the result of two components: 
an active radio galaxy, with jets and lobes, plus aged diffuse emission, 
which is not refurbished with new electrons at present. Our $Chandra$ data
show that the radio active nucleus of the BCG has extended X--ray emission, 
which we classify as a low--luminosity corona.
A residual image of the $XMM$--$Newton$  brightness distribution shows the 
presence of a spiral--like feature, which we interpret as the signature of gas 
sloshing. The presence 
of a sub--group is clear in the surface brightness residual map, and it is 
supported also by the $XMM$--$Newton$ temperature analysis.
The optical bidimensional analysis shows substructure in A\,3560. 
A galaxy clump has been found at the location of the X--ray sub--group, 
and another group is present south of the cluster core, in the region
where the spiral--like feature has been detected.
The aged part of the radio emission closely follows the spiral pattern of 
the X--ray residual brightness distribution, while the two active radio 
lobes are bent in a completely different direction. 
We conclude that the complex radio emission associated with the cluster BCG is 
the result of a minor merger event in A\,3560. The aged diffuse emission is 
strongly affected by the sloshing motion in the intracluster gas, and most 
likely bears information on the trajectory of 
the cluster BCG during the dynamical evolution of the cluster. On the
other hand, the bent jets and lobes of the current radio AGN activity may
reflect a complex gas velocity field in the innermost cluster regions and/or
sloshing--induced oscillations in the motion of the cD galaxy.}
{}
\maketitle
\section{Introduction}\label{sec:intro}

The central regions of galaxy clusters are sites of wonder. The bright 
magnitude of the central dominant galaxies (often showing multiple nuclei), 
the distribution of the intracluster medium (ICM), visible through thermal 
Bremsshtrahlung emission in the X--ray band, the morphology of the 
non--thermal radio emission associated with the central galaxy and other 
cluster members, and, in a number of cases, the diffuse radio emission coming 
from the ICM itself, are the observational signatures of the processes 
leading to formation of clusters, and bear invaluable information on their 
dynamical state.

The brightest cluster galaxies (BCGs), the most massive and luminous elliptical
galaxies in the Universe, show extreme broad band properties, most likely
related to their formation history, i.e. galaxy mergers and accretion in a 
hierarchical scenario of cluster formation (Lin \& Mohr \cite{lin04}, Bernardi
et al. \cite{bernardi06}). They are usually active in the radio band: 
the fraction of radio loud BCGs is much higher than for the other elliptical 
galaxies, reaching up to 30\% for L$_{\rm 1.4~GHz}>10^{23}$ W Hz$^{-1}$, and their 
radio luminosity function is the highest among elliptical galaxies 
(Best et al. \cite{best07}). 
The radio loud BCGs are usually transition objects between Fanaroff--Riley 
(Fanaroff \& Riley \cite{fr74}) type I and II radio galaxies (Owen \& Laing 
\cite{owen89}), and in many cases their morphology clearly reflects a strong 
interaction between the radio plasma and the dense ICM in the central cluster 
regions: the radio jets and lobes are often distorted and bent in a wide--angle 
shape, most likely as a consequence of a number of mechanisms, such as 
``cluster weather''(Burns \cite{burns98}; Mendygral et al. \cite{mendygral12}),
bulk motions in the ICM induced by the dynamical accretion of clusters, galaxy 
motion with respect to the ICM.

It has been recently proposed that merger--induced gas sloshing in the central 
regions of galaxy clusters may be responsible for the bent morphology of the 
central radio galaxy (Clarke et al. \cite{clarke04}; Ascasibar \& Markevitch 
\cite{ascasibar06}; Paterno--Mahler et al. \cite{pm13}). 
This mechanism may also provide the turbulence needed to 
re--accelerate the relativistic electrons responsible for the emission in 
``mini--halos'' (Giacintucci \& Mazzotta \cite{giacintucci08}; ZuHone et al. 
\cite{zuhone13}; Giacintucci et al. in prep.), diffuse  steep spectrum 
synchrotron radio sources with linear size of the order of \ltsim~500 kpc 
which are found in an increasing number of dynamically relaxed clusters 
(Giacintucci et al. \cite{giacintucci11a} and references therein).

With the advent of the {\it Chandra} and {\it XMM--Newton} X--ray 
observatories, over the past decade it has become clear that the interplay 
between the radio emission and the ICM in the central regions of galaxy 
clusters and groups may take the form of mechanical feedback to the ICM to 
prevent gas cooling (see McNamara \& Nulsen \cite{mcnamara07} and 
\cite{mcnamara12}). 
Indeed, a large fraction of massive galaxy clusters and groups have cooling 
regions centered on the radio loud cluster dominant galaxy, and a 
direct observational link between the active galactic central nucleus and 
the cooling ICM has been clearly established (i.e., Sun \cite{sun09}; Birzan et
al. \cite{birzan08}; Mittal et al. \cite{mittal09}; Birzan et al. 
\cite{birzan12}). 
Combined radio/X--ray studies show several impressive examples of cavities 
in the X--ray brightness distribution filled by radio lobes in rich clusters 
(some examples are Perseus, 
Fabian et al. \cite{fabian11} and references therein; MS\,0735+7421, 
Gitti et al. \cite{gitti07} and McNamara \& Nulsen \cite{mcnamara12}; Hydra A, 
Gitti et al. \cite{gitti11}) as well as in poor groups (i.e. HCG\,62, 
Gitti et al. 
\cite{gitti10}; NGC\,5044, David et al. \cite{david11}; the ``group survey''
performed with the Giant Metrewave Radio Telescope, GMRT, Giacintucci et al. 
\cite{giacintucci11b}). Observational support to the AGN/ICM feedback is 
finally provided by the evidence for recurrent bursts of radio activity 
from the central galaxy (e.g. NGC\,1407, Giacintucci et al. 
\cite{giacintucci12}) resulting in the presence of multiple cavities 
filled with radio emitting relativistc plasma (i.e. A\,262, Clarke et al. 
\cite{clarke09};  NGC\,5044, David et al. \cite{david11}; NGC\,5813, 
Randall et al. \cite{randall11}).

\medskip
In the context of the interaction between the central radio galaxy and the
intracluster medium in galaxy clusters, A\,3560 represents an intriguing case.
The cluster (z=0.048, richness class 3, 
L$_{\rm X~[0.2-10~keV]}=7.1\times10^{43}~$h$^{-2}$erg s$^{-1}$, 
$k$T=3.8 keV\footnote{We adopt a 
$\Lambda$CDM cosmology, with $\Omega_o=0.73, \Omega_{\rm M}=0.27$, 
H$_0$=71 km s$^{-1}$Mpc$^{-1}$. At the distance of A\,3560 1$^{\prime\prime}$=0.928
kpc.}, Bardelli et al. \cite{bardelli02}, hereinafter B02)  is located 
$\sim 1^{\circ}$ (corresponding to a projected distance of $\sim$3.3 Mpc) south 
of the A\,3558 cluster complex, the most massive structure in the Shapley 
Concentration, an exceptional region in the southern sky where cluster 
mergers and group accretion are still ongoing at the present cosmological 
epoch (Bardelli et al. \cite{bardelli98a} and \cite{bardelli98b}).
\\
Beppo--SAX observations of A\,3560 show that the ICM distribution is 
drop--shaped on the large scale, with a tail pointing north, in the direction
of the A\,3558 complex (B02). The X--ray brightness peak is off--centered 
from the centroid of the large scale gas distribution, and is coincident with 
the position of the BCG.
\\
The cluster radio emission is dominated by the powerful radio galaxy 
J\,1332--3308 (RA$_{\rm J2000}=13^{h}32^{m}25.8^{s}$, 
Dec$_{\rm J2000}=-33^{\circ}08^{\prime}09.4^{\prime\prime}$,
P$_{\rm 1.4~GHz}= 5.29\times10^{24}~{\rm W~Hz^{-1}}$),
associated with one of the two nuclei of the dumb--bell BCG. 
1.4 GHz and 5 GHz images at arcsecond resolution (B02), obtained with the 
Very Large Array (VLA, Socorro, New Mexico), clearly show a complex radio 
emission, as it is often found at the centre of galaxy clusters. Beyond the 
activity associated with the north--eastern component of the dumb--bell 
system, consisting of a  core and two extended lobes, other features are 
visible in those images, whose connection with the nuclear activity is not 
obvious. B02 concluded that the unusual and complex morphology of 
J\,1332--3308 could be either due to cluster weather, or the result of 
recurrent activity in the BCG.

Here we present a detailed multiwavelength study of J\,1332--3308 and of
the properties of the ICM in A\,3560, performed with the aim to understand 
the origin of the various components of the radio emission. 
Our analysis is based on data taken with the GMRT, VLA and Australia Telescope 
Compact Array (ATCA) over the frequency range 240 MHz -- 8.46 GHz, on 
{\it XMM--Newton} and {\it Chandra} proprietary data, and on optical data 
available in the literature. The paper is organised as follows: 
in Sect. 2 we describe the radio and X--ray observations and the data 
analysis;
the radio images and radio spectral study are presented in Sect. 3 and 4
respectively; in Sect. 5 we analyze the {\it XMM-Newton} and 
{\it Chandra} X--ray data, we examine the optical information from the 
literature, and perform a combined radio, X--ray and optical study of the 
cluster. 
Our results are discussed and interpreted in Sect. 6, and conclusions 
are given in Sect. 7. 
Throughout the paper we will use the convention S~$\propto\nu^{-\alpha}$.


\begin{table*}[h!]
\caption[]{Logs of the observations}
\begin{center}
\begin{tabular}{lccrrrc}
\hline
\hline\noalign{\smallskip}
Project ID & Array & Observing Date & $\nu$ & $\Delta\nu$ & Time & u--v range\\ 
           &       &                &   MHz &     MHz     & hour & k$\lambda$ \\
\hline\noalign{\smallskip}
14TVa01  & GMRT    & 12-08-08 &  240  &   8  &  8$^{\rm a}$    & 0.05 -- 20  \\
14TVa01  & GMRT    & 12-08-08 &  610  &  32  &  8$^{\rm a}$    & 0.2 -- 50   \\
14TVa01  & GMRT    & 16-08-08 & 1280  &  32  &  8$^{\rm a}$    & 0.4 -- 90   \\
AV234    & VLA-BnA & 20-06-98 & 1425  & 100  &  0.25$^{\rm b}$ & 1.5 -- 60   \\
AV234    & VLA-BnA & 20-06-98 & 4860  & 100  &  0.25$^{\rm b}$ & 5 -- 200    \\
AV246    & VLA-CnB & 13-03-00 & 8460  & 100  &  0.25$^{\rm b}$ & 2 -- 100    \\
hy13$^{\rm c}$& ATCA & 17-05-94 & 2366 & 128  &   2.5$^{\rm b}$  & 0.5 -- 50   \\
\hline\noalign{\smallskip}
\end{tabular}
\end{center}
\label{tab:logs}
Notes: $^{\rm a}$ Full track observations. The time reported refers to the 
total duration of the observation. $^{\rm b}$ Snapshot observations. The time 
reported refers to the time on--source. $^{\rm c}$ Archival data published in
Reid et al. (\cite{reid98}) and in Venturi et al. (\cite{venturi00}).
\end{table*}


\section{Observations and data reduction}\label{sec:obs}

In order to study the spectrum of the radio emission of J\,1332--3308 with 
at least 5 data points, and possibly reveal emission over a scale larger than 
that imaged in B02, we performed observations over a wide range of frequencies 
and angular resolutions. In particular, we observed J\,1332--3308 at 240 MHz, 
610 MHz and 1.28 GHz with the GMRT; at 1.4 GHz, 4.86 GHz and 8.46 GHz with 
the VLA. To complete the spectral coverage of our analysis we made use
of ATCA archival data at 2.36 GHz. Table 1 provides the details of the 
observations.

\subsection{The GMRT observations}

The 1.28 GHz and 610 MHz observations were performed using both the upper 
and lower side bands. The observations at 610 MHz and 240 MHz were carried
out using the dual receiver 240/610 MHz. The total observing bandwidth
was 32 MHz at 1.28 GHz and at 610 MHz, and 8 MHz at 240 MHz. 
\\
The data were collected in spectral--line mode at all frequencies, i.e. 256 
channels at 1.28 GHz and 610 MHz, and 64 channels at 240 MHz, with a spectral 
resolution of 125 kHz/channel. The calibration and data reduction were 
performed using the NRAO Astronomical Image Processing System (AIPS) package. 
An accurate editing was needed to identify and remove the data affected by 
radio frequency interference (RFI) at 240 MHz. In order to find a compromise 
between the size of the dataset and the need to minimize bandwidth smearing 
effects within the primary beam, after bandpass calibration the 
channels in each individual dataset were averaged to 6 channels of $\sim$1 
MHz each at 240 MHz, and $\sim$2 MHz each at 610 MHz and 1.28 GHz. 

The field of view around A\,3560 includes the strong source J\,133335--330523 
(PKS\,1330--328, RA$_{\rm J2000}=13^{h}33^{m}35^{s}$, 
Dec$_{\rm J2000}=-33^{\circ}05^{\prime}24^{\prime\prime}$), located at $\sim15^{\prime}$ 
east of the cluster centre,  which required multi--field calibration and 
imaging. Moreover, at 240 MHz special care was taken to remove the effects of 
the strong extended ($\sim 40^{\prime}$) radio galaxy PKS\,1333--33 
(Killeen et al. \cite{killeen86}), located at $\sim1.2^{\circ}$ south--east of 
the cluster centre\footnote{The total flux density of PKS\,1333--33, after
correction for the primary beam attenuation, is S$_{\rm 240~MHz}=39.5\pm2.8$ Jy,
in very good agreement with the total spectrum obtained using the total
intensity information available in the literature and in the data archives.}.
\\
After a number of phase self-calibration cycles, we produced the final 
images at each frequency. At 1.28 GHz and 610 MHz the upper and lower side 
band datasets were self-calibrated separately and then combined to make 
the final images. In order to compare the emission at different frequencies
and on different angular scales we produced full resolution and tapered images, 
covering a resolution range from $\sim5^{\prime\prime}\times2^{\prime\prime}$
to $\sim20^{\prime\prime}\times12^{\prime\prime}$. 

The rms noise level (1$\sigma$) achieved in the final full resolution images 
is $\sim 0.1$ mJy~b$^{-1}$ at 1.28 GHz, $\sim 0.3$ mJy~b$^{-1}$ at 
610 MHz and $\sim 2$ mJy~b$^{-1}$ at 240 MHz (see Table 2). 
The average residual amplitude errors in our data are of the 
order of $\la$5\% at 1.28 GHz and 610 MHz, and of the order of 8\% at 
240 MHz. The most relevant parameters of the final full resolution
images are reported in Table \ref{tab:images}.

\subsection{The VLA observations}

J\,1332-3308 was observed with the VLA in the hybrid BnA configuration at 
1.4 and 4.86 GHz, and with the CnB configuration at 8.46 GHz 
(see Table \ref{tab:logs}) as part of a project devoted to the study the 
nuclear activity of the extended radio galaxies in the central cluster 
complexes of the Shapley Concentration (Venturi et al. in prep.). 
At all frequencies the observing bandwidth was 100 MHz.
Following a standard approach, the data were edited, self--calibrated
and imaged using AIPS. 
The 1$\sigma$ rms ranges from $\sim 0.02$ mJy~b$^{-1}$ to 
$\sim 0.13$ mJy~b$^{-1}$ to from 4.86 GHz to 1.425 GHz respectively.
The residual amplitude errors are of the order of few \%.
The most relevant parameters of the final full resolution
images are reported in Table \ref{tab:images}.


\begin{table}[h!]
\caption[]{Characteristics of the final images of J\,1332--3308}
\begin{center}
\resizebox{0.5\textwidth}{!}{%
\begin{tabular}{cccrlc}
\hline
\hline\noalign{\smallskip}
Array &  $\nu$ & Resolution & P.A. & rms & S$_{tot}$\\ 
   &   MHz & $^{\prime\prime}\times^{\prime\prime}$ & $^{\circ}$  & mJy~b$^{-1}$ & mJy\\
\hline\noalign{\smallskip}
GMRT    &   240  &  18.6$\times$11.6 & --3 & $\sim$ 2.0  & 4850 $\pm$ 399\\
GMRT    &   610  &   8.2$\times$5.0  & 0   & $\sim$ 0.30 & 2678 $\pm$ 134\\
GMRT    &  1280  &   4.7$\times$2.1  & 25  & $\sim$ 0.10 & 1289 $\pm$ 63\\
VLA-BnA &  1425  &   4.2$\times$3.7  & 61  & $\sim$ 0.13 & 1021 $\pm$ 31\\
ATCA 	&  2366  &   9.4$\times$6.2  & 2   & $\sim$ 0.15 & 675 $\pm$ 34\\
VLA-BnA &  4860  &   1.7$\times$1.2  & 13 & $\sim$ 0.020 & $~$229 $\pm$ 12$^{(a)}$\\
VLA-CnB &  8460  &   2.7$\times$1.8  & 44 & $\sim$ 0.025 & 213 $\pm$ 11\\
\hline\noalign{\smallskip}
\end{tabular}}
\end{center}
\label{tab:images}
Notes: $^{(a)}$ This value is underestimated and should be considered as a
lower limit (see Sect. 3.1).
\end{table}


\subsection{X-ray observations and data reduction}
\label{sec:xray_obs}
We complemented the radio information on  J\,1332--3308 with two X-ray
observations on A\,3560 performed with {\it XMM--Newton}  and {\it
  Chandra}, in order to characterize the interaction between the
central radio galaxy and the ICM. Details on the observations are
provided in Table \ref{tab:xray_obs}.


\begin{table}[h!]
\caption[]{Characteristics of the X-ray observations of A\,3560}
\begin{center}
\resizebox{0.5\textwidth}{!}{%
\begin{tabular}{cccc}
\hline
\hline\noalign{\smallskip}
Instrument & Obs.ID & Detector &Exposure (ks)\\
\hline\noalign{\smallskip}
& &MOS1 & 29 \\
{\it XMM-Newton} & 0205450201 & MOS2 & 31 \\
 & & pn & 21 \\
{\it Chandra} &  12883 & ACIS-S3 & $18.7$ \\
\hline
\end{tabular}}
\end{center}
\label{tab:xray_obs}
\end{table}


\subsubsection{{\it XMM--Newton} data analysis}
\label{sec:xray_obs_xmm}
{\it XMM--Newton} observed A\,3560 in 2004 for a total exposure of 46
ks. 
We generated calibrated event files using the SAS software v.\,$12.0$
and  we removed soft proton flares using a double filtering
process, first in a hard (10--12 keV) and then in a soft (2--5 keV)
energy range. The resulting clean exposure times are 29, 31, and 21 ks
for MOS1, MOS2, and pn, respectively. The event files were filtered
according to {\verb PATTERN }  and {\verb FLAG } criteria. 
Bright point-like sources were detected using a procedure based on the
SAS task  {\verb edetect_chain }  and removed from the event files. 
As background files, we merged nine blank field observations,  we
reprojected them to match the source observation and renormalized by the
ratio of the count-rates in an external region and in the hard band to
account for possible temporal variations of the particle background
(more details in Leccardi et al. \cite{leccardi10}).
This procedure does not introduce substantial distortions in the soft
energy band, where the cosmic background components are more important,  
because we limited our analysis to regions where the source outshines 
the background in the soft band by more than one order of magnitude.\\
For the spectral analysis, we extracted spectra for the three EPIC
detectors, with their corresponding redistribution matrix file (RMF)
and effective area (ARF).  We performed spectral fitting within XSPEC
v$12.0$, jointly for the three detectors in the energy band $0.7-7$
keV. We used an  absorbed APEC model, leaving as free parameters the
ICM temperature, the metal abundance, and the normalization, while we
fixed the redshift ($z=0.048$) and $N_H$ to the galactic value
($4.18\times10^{20}\,\rm{cm}^{-2}$, Kalberla et al. \cite{kalberla})

\subsubsection{{\it Chandra} data analysis}
We obtained an exploratory observation pointed at the centre of
A\,3560 with {\it Chandra} in Cycle 12. The cluster was observed in
April 2012 for 20 ks with the ACIS-S instrument.
We reduced the observation with the software CIAO v$.4.4$ and the
calibration database (CALDB) version $4.4.8$. We reprocessed data
from level 1 event files, following the  standard {\it Chandra}
data-reduction threads\footnote{http://cxc.harvard.edu/ciao/threads/index.html}.
We applied the standard corrections to  account for a time-dependent drift 
in the detector gain and charge transfer inefficiency, as implemented in the 
CIAO tools. From low surface brightness regions of the active chips we 
extracted a light-curve ($5-10$ keV) to identify and excise periods of 
enhanced background. 
We detected point sources with the CIAO tool {\tt wavdetect}.

\section{The complex morphology of J\,1332--3308}\label{sec:img}

At all frequencies, the field of view of A\,3560 is dominated by
the radio galaxy J\,1332--3308, associated with the dominant dumb--bell
central cluster galaxy. The strong source J\,133335--330523 is a background 
object at redshift z$\sim$ 0.08 (see Section 2.1), which becomes relevant
at low frequencies, owing to its steep spectrum and less severe primary
beam attenuation.  
The remaining few radio sources in the field are background objects, 
with and without optical identification.

\subsection{The active core and the lobes}

Figure 1 shows the radio emission associated with the BCG in A\,3560 in
decreasing frequency and resolution order.
In the left panel we provide the radio emission of J\,1332--3308 at 4.86 GHz 
and 8.46 GHz (red and black contours respectively), imaged at comparable 
angular resolution, overlaid on the optical red DSS--2 frame.
At both frequencies the radio galaxy shows an active nucleus, two extended 
lobes and further emission south of the southern lobe, referred to as
``diffuse emission'' hereinafter.
The core of the radio emission is associated with the north--eastern
nucleus of the dumb--bell system. There is a clear radio emission gap between
the core and the lobes of the radio galaxy: only the southern jet becomes 
visible at  $\sim 15^{\prime\prime}$ (i.e. $\sim$ 14 kpc) from the core, to 
merge with the southern lobe. We did not
classify it as a jet in the 4.8 GHz image presented in B02, due to its worse 
quality, and referred to it as
``filament''. The set of images presented in this paper leaves no 
doubt on the origin of this feature, which we will refer to as ``southern jet''
hereinafter. 


\begin{figure*}[htbp]
\centering
\includegraphics[angle=0,scale=0.65]{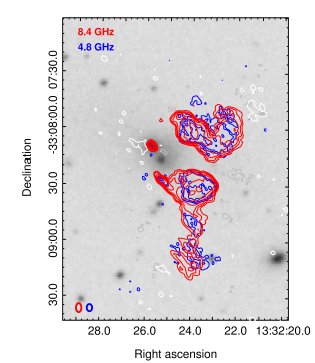}
\hspace{0.3cm}
\includegraphics[angle=0,scale=0.65]{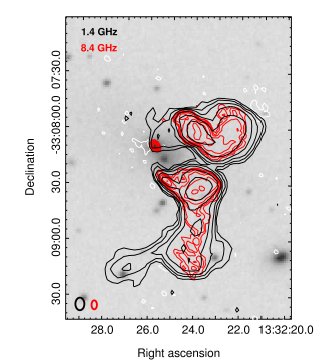}
\hspace{0.3cm}
\includegraphics[angle=0,scale=0.65]{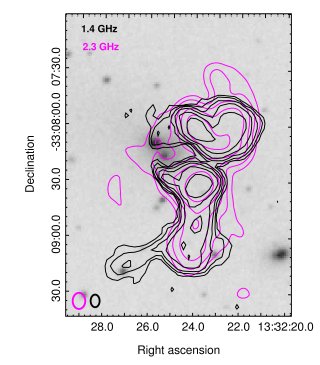}
\caption{{\it Left panel:} DSS--2 red optical image of J\,1332--3308 with 
overlaid VLA radio contours at 4.8 GHz, shown in blue (negative are grey)
and 8.4 GHz (shown in red, negative are white).
4.8 GHz radio contours are $\pm$0.075,0.3,1.2,4.8 mJy~b$^{-1}$, the 
resolution is $1.7^{\prime\prime}\times1.2^{\prime\prime}$, p.a. 13$^{\circ}$, 
and the 1$\sigma$ noise level is 0.020 mJy~b$^{-1}$. 
The 8.4 GHz contours are 
$\pm$0.08,0.16,0.32,0.64,1.28,2.56 mJy~b$^{-1}$, the resolution is 
$2.7^{\prime\prime}\times1.8^{\prime\prime}$, p.a. 44$^{\circ}$,
and the 1$\sigma$ noise level is 0.025 mJy b$^{-1}$.
{\it Central panel:} DSS--2 red optical image of J\,1332--3308 with 
overlaid VLA radio contours at 8.4 GHz, shown in red (same as left panel)
and 1.4 GHz (shown in black). The 1.4 GHz contours start 
at $\pm$0.39 mJy b$^{-1}$ (3$\sigma$) and are spaced by a factor of 2,
the resolution is $4.2^{\prime\prime}\times3.7^{\prime\prime}$, p.a. 61$^{\circ}$.
{\it Right panel:} DSS--2 red optical image of J\,1332--3308 with 
overlaid ATCA radio contours at 2.3 GHz (shown in magenta)
and 1.4 GHz VLA contours in black (same as central panel).
The 2.3 GHz contours are $\pm3\sigma=0.45$ mJy b$^{-1}$ (3$\sigma$)
and spaced by a factor of 2, the resolution is 
$9.4^{\prime \prime}\times 6.2^{\prime \prime}$,  p.a. 2$^{\circ}$.
In each panel the restoring beams are reported in the bottom left corner.}
\label{fig:fig1}
\end{figure*}
%

We note that the central u--v gap at 4.86 GHz (see Table 1) does not allow the 
detection of emission exceeding an angular size of $\sim 40^{\prime\prime}$. 
Since the extended emission is the dominant feature of J\,1332--3308 at
all frequencies, the 4.86 GHz flux density measured from our observations 
is considerably underestimated. For this reason we will not consider this 
frequency in the spectral analysis carried out in Sect. 4.1.

In the central panel of Fig. 1 we report the radio emission at 1.4 GHz 
and at 8.4 GHz (black and red contours respectively). 
This overlay shows further interesting features. At 1.4 GHz a brigde of 
emission connects the radio core with the northern lobe. 
This feature, clearly visible in  the central and right panel of Fig. 1 at 
1.4 GHz and 2.3 GHz, is most likely related to the extended radio emission 
rather than to a jet itself (see Sect. 3.2).
Moreover, the radio emission south of the southern lobe is considerably more 
extended than at the higher frequencies, and it is characterized by a
boxy brightness distribution. Finally, a thin ``spur'' is detected 
east of the southernmost part of the radio emission (see next subsection). 
All these features are labelled in the left panel of Fig. 2, for a 
proper comprehension.

The right panel of Fig. 1 shows the ATCA 2.3 GHz image (blue contours), with 
the VLA 1.4 GHz contours overlaid for comparison.
The ATCA observations were centered on the Shapley Concentration cluster 
A\,3562, located more than 1$^{\circ}$ north of A\,3560 (Venturi et al.
\cite{venturi00}), thus the emission from J\,1333--3308 suffers from severe 
primary beam attenuation, which reduces the signal--to--noise level in
the sky region we are dealing with.
A bridge of emission between the core and the lobes is visible in the
figure, as already noticed at 1.4 GHz.

%
\begin{figure*}[htbp]
\centering
\includegraphics[angle=0,scale=0.82]{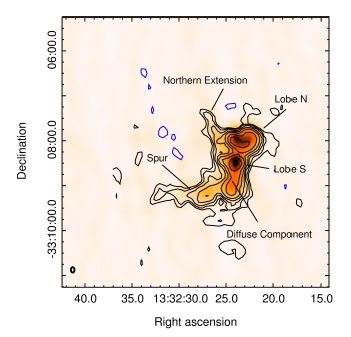}
\hspace{0.8cm}
\includegraphics[angle=0,scale=0.75]{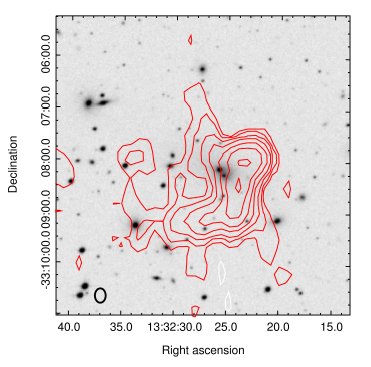}
\caption{{\it Left panel}: GMRT 610 MHz radio contours of  J\,1332--3308 at the 
resolution of $8.2^{\prime \prime}\times 5.0^{\prime \prime}$, p.a. 0$^{\circ}$ overlaid
on the GMRT 240 MHz colour image (same as right panel).
The 1$\sigma$ level at 610 MHz is $\sim$0.28 
mJy b$^{-1}$. Contours are spaced by a factor 2 starting from 
$\pm3\sigma=0.85$ mJy beam$^{-1}$. Negative contours are shown in blue.
{\it Right panel}: GMRT 240 MHz radio contours of J\,1332--3308 at the 
resolution of $18.6^{\prime \prime}\times 11.6^{\prime \prime}$, p.a. -3.2$^{\circ}$ 
overlaid on the DSS--2 image. The 1$\sigma$ noise level is $\sim$2 
mJy b$^{-1}$. Contours are spaced by a factor 2 starting from $\pm3\sigma=5.5$ 
mJy b$^{-1}$. Negative contours are shown in white.
In each panel the restoring beam is reported in the bottom left corner.}
\label{fig:fig2}
\end{figure*}
%

\subsection{Large scale radio emission}

The left panel of Fig. 2 shows an overlay of the 610 MHz (contours) and 
240 MHz GMRT radio emission (colour scale), while the right panel reports
the radio contours at 240 MHz overlaid on the red DSS--2 plate.
It is clear that at 610 MHz and 240 MHz the extent of
J\,1332--3308 is considerably larger than at frequencies around and above 
1.4 GHz, especially eastwards. 
Beyond the core, the lobes, the southern jet and the diffuse component,
visible  at all frequencies, we further image two filamentary structures: 
the spur, already visible at 1.4 GHz, is considerably more extended 
here, and the northern extension, located north--east of the core. 
These are labelled in the left panel of Fig. 2.
The 240 MHz contours in the right panel of Fig. 2 clearly show that that 
the spur and the northern extension seem to form a portion of a ring, east
of the location of the BCG. 

The lenght of the spur considerably increases with decreasing frequency, 
going from $\sim$ 28 kpc at 1.4 GHz, to $\sim$ 90 kpc at 240 MHz. 
Hints of the ''northern extension'' are visible at 1.4 GHz, too, however only 
the low frequency images support the reliability of this feature (left 
panel of Fig. 2). 
Assuming that even these faint features belong to J\,1332--3308, the total 
extent of the radio galaxy at 240 MHz is $\sim4^{\prime}$, i.e. $\sim$ 220 kpc.

Our set of images clearly shows that the radio emission associated with the BCG 
at the centre of A\,3560 is very complex. In particular, the connection between 
the spur, the northern extension and the diffuse component with the other AGN 
features of J\,1332--3308 (radio core and lobes) is not trivial, and
may hide a rather complex physical origin of the whole radio emission.

\section{Spectral analysis}\label{sec:spx}

In order to understand the nature of the various components of the complex 
radio emission of J\,1332--3308,  we used the datasets presented here to 
perform a spectral study of this source in the frequency range 240 MHz -- 8.46 
GHz, by means of the information at 6 different frequencies. As 
clarified in Sect. 3.1, the 4.86 GHz data are not used in this analysis.  
In the previous section we did not show the 1.28 GHz GMRT images, 
since they do not add any information to the 1.4 GHz image.
However, we used the flux density information 
at this frequency to perform our spectral analysis.
\\
The total spectrum of the source and the integrated spectra of its components 
are discussed in subsection 4.1. The spectral index images of the source are
presented in subsection 4.2. 

\subsection{Integrated spectra}

The total integrated spectrum of J\,1332--3308 was derived using the
flux density measurements reported in the last column of Table 2. 
To those values we
added the flux density value at 150 MHz derived from the TGSS
(TIFR GMRT Sky Survey{\footnote{http://tgss.ncra.tifr.res.in/}}), 
which has a resolution of $24^{\prime \prime}\times 15^{\prime \prime}$, 
p.a. 30$^{\circ}$, and rms ($1\sigma$) $\sim$ 11 mJy b$^{-1}$ in the A\,3560 
field region. 
The total flux density of J\,1332--3308 at this frequency is 
S$_{\rm 150 MHz}=9.3\pm1.9$ Jy.
The flux density measurements at each frequency were obtained by means of 
the AIPS task TVSTAT integrating within the 3$\sigma$ contour level at each 
frequency. The total spectrum is reported in Fig. 3 (black), with  
the spectral fit overlaid (see Sect. 4.2).
\\
The total flux density of J\,1332--3308 is dominated by the extended 
components, and the resulting spectrum is steep, with 
$\alpha^{\rm 8.46 GHZ}_{\rm 150 MZ}=0.91\pm0.02$, if fitted with a single power--law.
The total flux density values of J\,1332--3308 reported in Table 1 are
slightly higher than the sum of the individual components at each observing 
frequency, the difference increasing from $\sim$5\% to $\sim$10\% with
decreasing frequency. This highlights the presence of diffuse emission, 
expecially at frequencies $\nu\le$ 1.4 GHz.


\begin{figure}[htbp]
\centering
\includegraphics[angle=0,scale=0.45]{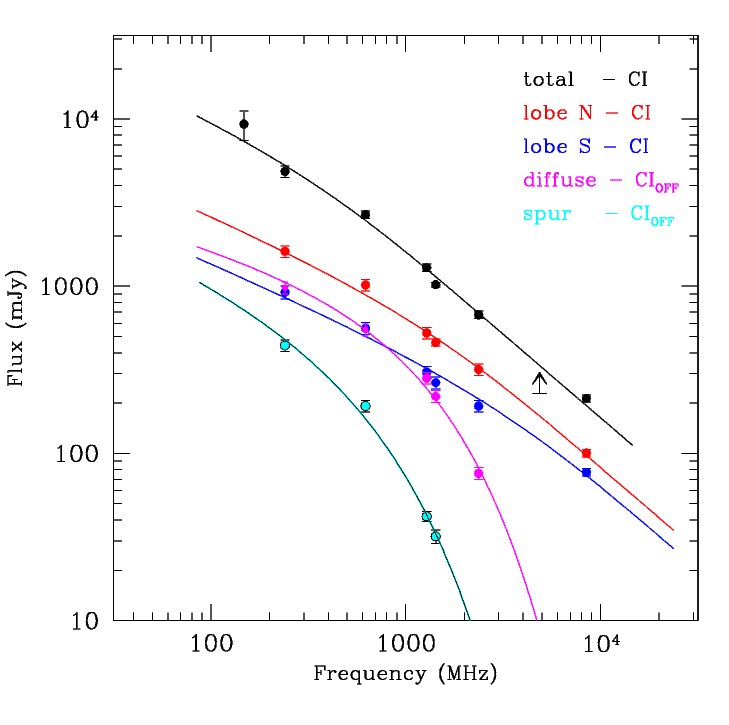}
\caption{Integrated spectra of the total radio emission and individual
components of J\,1332--3308. The black arrow represents the lower limit
to the total flux density at 4.86 GHz.}
\label{fig:fig3}
\end{figure}


The study of the integrated spectrum of each individual component of the 
radio emission of J\,1332--3308 required further imaging at resolutions
different than those reported in Figs. 1 and 2, so as to derive flux 
density values at comparable resolutions.
The VLA and ATCA images used for the integrated spectra are those presented 
in Table 2. At 610 MHz we produced a tapered image  with a resolution of 
$13.9^{\prime \prime}\times 6.8^{\prime \prime}$, p.a. -17.9$^{\circ}$, achieving a 
1$\sigma$ rms $\sim$ 0.55 mJy b$^{-1}$ in order to detect all the diffuse 
emission (spur and diffuse component).
The core emission of J\,1332--3308 can be isolated only at 8.4 and 4.8 GHz 
(see Fig. 1), but a consistent estimate of its flux density was made 
also at 2.3 GHz and at 1.4 GHz, searching for the peak of the emission 
in the core area using the task IMFIT. 
No estimate of the core flux density was possible at frequencies below
1.4 GHz.
\\
As pointed out in Sect. 3.2, and clear from Figs. 1 and 2, the size of 
the spur  increases considerably at decreasing frequency. In order 
to estimate its total spectrum we integrated the 1.4 GHz and 610 MHz flux 
density over the same region of the 240 MHz emission. The flux density
values of each individual component are reported in Table 3. We point out
that the uncertainties associated with each measurement take into account
the difficulty in isolating the individual components, especially at
low frequency and at resolutions lower than $\sim 5^{\prime\prime}$. 

%
\begin{table}[h!]
\caption[]{Flux density and spectral index for the components of J\,1332--3308.}
\begin{center}
\begin{tabular}{lrc}
\hline
\hline\noalign{\smallskip}
Component &  $\nu$ & S$\pm\Delta$S \\ 
          &    MHz &      mJy\\
\hline\noalign{\smallskip}
Lobe N    &   8460 &  100$\pm$5 \\ 
          &   4860 &  ~119$\pm$6$^{(a)}$  \\
          &   2366 &  318$\pm$25  \\
	  &   1425 &  460$\pm$23 \\
          &   1280 &  525$\pm$42 \\
	  &    610 & 1017$\pm$80 \\
	  &    240 & 1614$\pm$129 \\
Lobe S    &   8460 &   77$\pm$4 \\
          &   4860 & ~86$\pm$4$^{(a)}$ \\
	  &   2366 &  192$\pm$15 \\
	  &   1425 &  265$\pm$21 \\
	  &   1280 &  306$\pm$25 \\
	  &    610 &  560$\pm$45 \\
	  &    240 & 1009$\pm$81 \\
Jet S     &   8460 &  7.0$\pm$0.2  \\
          &   4860 &  ~4.6$\pm$0.3$^{(a)}$ \\
          &   2366 &   17$\pm$2  \\
	  &   1425 &   26$\pm$2 \\
	  &   1280 &   31$\pm$3 \\
	  &    610 &   69$\pm$6\\
Diffuse comp. &   8460 & 17$\pm$1 \\
              &   4860 & ~16$\pm$1$^{(a)}$ \\
	      &   2366 & 76$\pm$6 \\
	      &   1425 & 219$\pm$18  \\
	      &   1280 & 282$\pm$22 \\
	      &    610 & 550$\pm$44 \\
	      &    240 & 978$\pm$78 \\
Core &   8460 &    5.7$\pm$0.2   \\
     &   4860 &    ~3.9$\pm$0.3$^{(a)}$  \\
     &   2366 &    3.7$\pm$0.2    \\
     &   1425 &    1.9$\pm$0.2    \\
Spur &   1425 &    32$\pm$3   \\
     &   1280 &    42$\pm$3   \\
     &    610 &   192$\pm$15  \\
     &    240 &   442$\pm$35  \\
\hline\noalign{\smallskip}
\end{tabular}
\end{center}
\label{tab:fluxcomp}
Notes: $^{(a)}$ See text in Section 3.1.
\end{table}

The spectra of the northern and southern lobes (Fig. 3, red and
blue respectively) are similar in shape, steepness and normalization, 
indicating a common origin and evolution.
The northern lobe has an average spectral index 
$\alpha^{8.46 GHz}_{610 MHz}$ = 0.88$\pm$0.05, slightly steeper than  that 
of the southern one, $\alpha^{8.46 GHz}_{610 MHz}$ = 0.75$\pm0.05$.
The spectra of both lobes show a flattening at $\nu \leq$ 610 MHz, with
$\alpha^{610 MHz}_{240 MHz}$ = 0.50$\pm$0.13 and  
$\alpha^{610 MHz}_{240 MHz}$ = 0.63$\pm$0.13  for the northern and southern lobe 
respectively. Due to the large uncertainties associated with the low 
frequency flux density measurements, these values agree within
1$\sigma$ (see Table 3). 
\\
The spectral index of the southern jet (not shown) is 
$\alpha^{8.46 GHz}_{610 MHz}$ = 0.87$\pm$0.06 (see Table 3), a typical value for 
arcsecond scale radio jets. The core shows an inverted spectrum, with
$\alpha_{\rm 1.4~GHz}^{\rm 8.46~GHz}=-0.62\pm0.08$, as 
expected for a compact component with optically thick emission. The radio 
nucleus is thus currently active.

The spectra of the diffuse component and of the spur (Fig. 3, magenta and
cyan respectively) differ considerably from the lobes, suggesting a 
different evolutionary stage and/or a different origin of the radio emitting 
plasma. In both cases a pronounced steepening at frequencies $\ge~610$ MHz 
is clear: in the frequency range 240--610 MHz the spectral index is 
0.62$\pm$0.16 and 0.89$\pm$0.16 respectively for the diffuse components and 
the spur, while at higher frequencies we have respectively 
$\alpha_{\rm 1.28~GHz}^{\rm 2.3~GHz}=2.13\pm0.26$, and 
$\alpha_{\rm 610~MHz}^{\rm 1.28~GHz}=2.05\pm0.20$. 
  
\subsection{Spectral fit and radiative ages}

The integrated spectra of the various components of emission of
J\,1332--3308 provide invaluable information on their radiative ages,
which we estimated by fitting our data using the Synage++ package 
(Murgia \cite{murgia01}). The total spectrum of the source and the curved 
spectra of the lobes {\bf (Fig. 3)} are best fitted by a continuous injection 
model (CI, Kardashev \cite{kardashev62}), in which the radio source is 
continuously refurbished by a flow of relativistic particles with a
power--law energy distribution in a region of constant magnetic field. 
In all cases we assumed a fixed injection spectral index 
$\alpha_{\rm inj}=0.5$, which provides the best fit. The best fit values 
for $\nu_{\rm br}$ are reported in Tab. 4.

The major steepening of the diffuse component and of the spur for 
$\nu~$\gtsim~1 GHz {\bf (Fig. 3)} require a composite fit, with injection 
followed by a nuclear switch--off of the supply of freshly accelerated 
particles, and a ``relic'' phase (CI$_{\rm OFF}$ model, Murgia et al. 
\cite{murgia11}). 
The best fit is provided assuming a fixed injection spectral index 
$\alpha_{\rm inj}=0.5$. The best fit values for $\nu_{\rm br}$ are 
reported in Tab. 4.

In both cases we obtained t$_{\rm off}$/t$_{\rm rad}\sim 1$, where 
t$_{\rm off}$/t$_{\rm rad}$ is the ratio between the dying phase and the age 
of the component. This result suggests that for both components the
injection of fresh electrons has ceased long ago, and the spectrum is
now dominated by radiative losses.

We estimated the radiative age of all components under the assumption of
equipartition (see Giacintucci et al. \cite{giacintucci12} for details)
and using the break frequency $\nu_{\rm br}$ derived from the spectral fit 
by means of the relation:

$$t{\rm (Myr)}=1590[\nu_{\rm br}(1+z)]^{\rm -1/2}\times({\rm B_{eq}^{0.5}/(B_{eq}^{\rm 2}+B_{CMB}^{\rm 2}}))$$

where B$_{\rm CMB}$=3.25(1+z)$^2$. Our results are reported in Table 4.

For the lobes, the equipartition values for the magnetic field and internal 
pressure are typical of active extended radio galaxies. The derived
radiative ages point to a very new cycle of activity, consistent with
the inverted spectrum of the core. If we assume that the source is in the
plane of the sky, and take into account the bent morphology of the jets, 
we obtain a plausible value for the advance speed of the two lobes, 
i.e. v$_{\rm adv}\sim$0.06c, consistent with previous studies 
(e.g., Myers \& Spangler \cite{myers85}; Scheuer \cite{scheuer95}).

The estimated radiative ages for the diffuse component and the spur deserve 
a comment. They are of the order of $\sim$ 100--200 Myr, i.e. two orders of 
magniture older than the lobes. These values should be taken with care, 
since in both cases considerable mixing with the ICM might have already taken 
place, however they are very similar to what is found both for dying radio 
galaxies in clusters (Murgia et al. \cite{murgia11}), and for the aged 
emission in restarted radio galaxies (e.g. NGC\,1407, Giacintucci et al. 
\cite{giacintucci12}). 

We wish to add a final comment of the total spectrum of J\,1332--3308
(Fig. 3). Its shape and spectral fit are typical of active radio galaxies, 
since the flux density of the lobes dominates at all frequencies (as
clear from Tab. 3), however the analysis performed on the individual
components clearly shows that the total radio emission is actually formed 
by a blend of an active radio galaxy and aged plasma. Any conclusion 
drawn from the total spectrum itself would thus suffer from serious
biases.

\subsection{Spectral index imaging}
We imaged the spectral index distribution of the radio emission in 
J\,1332--3308 in two frequency ranges:
the distribution of $\alpha_{1.4~GHz}^{8.4~GHz}$ 
provides information on the lobes and on the diffuse component; the 
distribution of $\alpha_{240~MHz}^{610~MHz}$ allows us to address the nature 
of the filamentary features (the spur and the northern extension) visible 
at low frequencies (see Fig. 2), best suited to trace aged relativistic plasma.
\\
In order to ensure a comparable u--v coverage in the images used to produce 
the spectral index distribution in the two frequency ranges, we used the 
data in the common portion of 
the u--v plane of the respective datasets. As a consequence, the quality
of the individual images used to derive the spectral index distribution
is slightly worse than that of the final images presented in Section 3.
To produce the spectral index image we clipped the images at each 
frequency for pixel values below the $3\sigma$ level.
The resolution of the $\alpha_{1.4~GHz}^{8.4~GHz}$ image is 
$4.5^{\prime\prime}\times4.0^{\prime\prime}$, while that of $\alpha_{240~MHz}^{610~MHz}$
is $18^{\prime\prime}\times12^{\prime\prime}$. 
Our results are shown in Fig. 4, with total intensity contours overlaid.


\begin{figure*}[htbp]
\centering
\includegraphics[angle=0,scale=0.47]{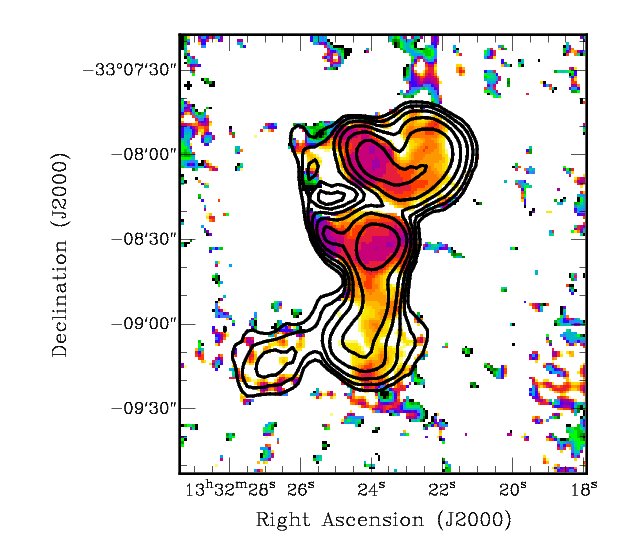}
\hspace{0.3cm}
\includegraphics[angle=0,scale=0.45]{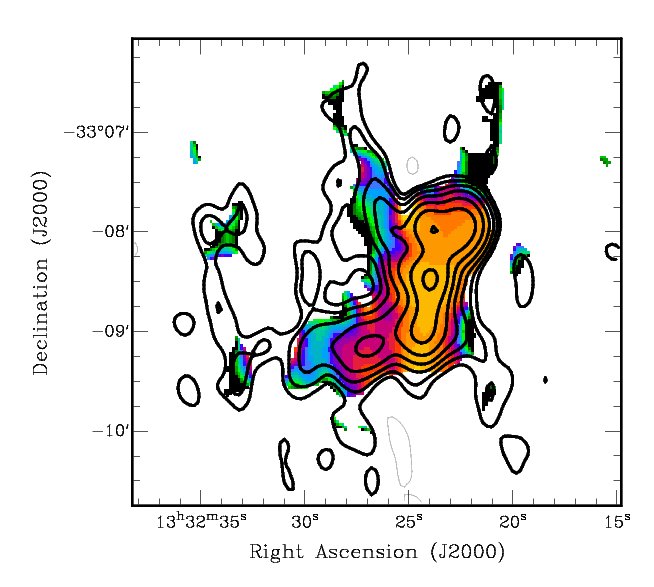}
\caption{{\it Left panel:} Spectral index distribution in J\,1332--3308 in the 
frequency range 1.4 GHz--8.4 GHz (colour scale) at the resolution of 
$4.5^{\prime\prime}\times4.0^{\prime\prime}$, with 1.4 GHz VLA contours overlaid 
(resolution $8.9^{\prime\prime}\times4.6^{\prime\prime}$, contours start at 0.84
mJy b$^{-1}$ and scale by a factor of 2). The colour scale is as follows: 
purple 0.6; orange 1.1; light orange 1.4.
{\it Right panel:} Spectral index distribution in J\,1332--3308 in the 
frequency range 240--610 MHz (colour scale) at the resolution of 
$18^{\prime\prime}\times12^{\prime\prime}$, with 240 MHz GMRT contours overlaid 
(same as Fig. 2 right panel). The colour scale is as follows: orange 0.5-0.6;
purple 1; blue 1.8.}
\label{fig:fig4}
\end{figure*}


The 1.4--8.4 GHz spectral image (left panel of Fig. 4) clearly shows that 
the lobes steepen away from the core of the radio emission, with the 
spectral index $\alpha_{1.4~GHz}^{8.4~GHz}$ in the range $\sim$0.6 (purple) to
1.1 (orange) going from the highest to the lowest surface brightness 
regions respectively. 
This image shows a continuity of the spectral index distribution from the 
jet to the southern lobe, consistent with the idea that the jet is refurbishing
the lobe with fresh electrons.
The spectral index shows substructure along the diffuse component: 
$\alpha_{1.4~GHz}^{8.4~GHz}\sim 1.2-1.4$ in the central ridge (orange), then it 
steepens out to $\sim 2$ (yellow) going outwards. This trend reflects the 
sharp boxy surface brightness  distribution of this component at all 
frequencies. 

The 240--610 MHz spectral index image (right panel of Fig. 4) 
clearly shows that the radio emission in J\,1332--3308 consists of two separate 
parts: the lobes and the diffuse component form a flat plateau, with values 
$\sim 0.5-0.6$ (orange), where electrons are still characterized by their
original power--law distribution, without significant evolution due to
ageing. On the other hand, the filamentary eastern features of the radio 
emission,  i.e. the spur and the northern extension, are considerably steeper.
The spectral index of the spur steepens from $\sim$ 1 (purple)
to 1.8 (blue) going eastwards.
The northern extension is the steepest feature, with $\alpha \sim 1.8$.
No transverse gradients are visibile, and the spur and northern extension
seem to trace a portion of a ring of  steep radio emission. 
\\
We point out the consistency between the spectral index values in the 
imaging and in the integrated spectra of each individual component in 
J\,1332--3308.

%
%
\begin{table}[h!]
\caption[]{Results from the spectral fitting.}
\begin{center}
\begin{tabular}{lccccc}
\hline
\hline\noalign{\smallskip}
Component & P$_{\rm 240~MHz}$ & $\nu_{\rm br}$ & B$_{\rm eq}$ & P$_{\rm min}$ & 
t$_{\rm rad}$ \\
          &  W Hz$^{-1}$      &     MHz      &   $\mu$G  & dyne cm$^{\rm -2}$ &  
Myr    \\
\hline\noalign{\smallskip}
Lobe N      & 8.4$\times10^{24}$ & $\sim$1780 &  5.8  & 1.9$\times10^{-12}$ & 
$\sim$1.9 \\
Lobe S      & 5.2$\times10^{24}$ & $\sim$4100  &  7.2 & 2.3$\times10^{-12}$ & 
$\sim$1.0 \\
Diff. Comp. & 5.1$\times10^{24}$ & $\sim$300   &  6.0 & 2.1$\times10^{-12}$ & 
$\sim$130 \\
Spur        & 2.3$\times10^{24}$ & $\sim$300   &  5.0 & 1.3$\times10^{-12}$ & 
$\sim$170 \\
\hline\noalign{\smallskip}
\end{tabular}
\end{center}
\label{tab:ages}
\end{table}

\section{X--ray and optical properties of A\,3560}

To understand the origin of the unusual radio emission at the centre 
of A\,3560 and investigate a possible connection with the ICM, we made 
use of proprietary X--ray data and optical literature information to 
study the thermal properties and dynamical state of the  cluster. 
Details on the X--ray observations and on data reduction are provided 
in Sec. \ref{sec:xray_obs}.

\subsection{ICM structure of A3560}
\label{sec:icm}

In the left panel of Fig. 5 we show the {\it XMM--Newton} image obtained 
from the combination of MOS1, MOS2 and pn cameras, as described in Rossetti 
et al. (\cite{rossetti07}). 
The X--ray surface brightness is asymmetryic, with a 
north--western elongation in the direction of the A\,3558 Shapley cluster 
complex, with a positional shift between the centre of the inner and outer
isophotes, suggestive of a disturbed dynamical state.
We used this image to produce a residual image of the surface brightness
distribution. We masked out point sources and calculated the average 
surface brightness, making different assumptions on the centroid of the 
model and on the geometry of the emission. In particular, we approximated 
the X--ray isophothes with concentric circular or elliptical annuli, 
centered either on the peak or on the centroid of the X--ray emission.
In each case we then subtracted this symmetric model from the image and 
divided the resulting image by the model. This approach allows us to 
provide a quantitative information on the relative asymmetric variations 
of the source.

\begin{figure*}[htbp]
\centering
\includegraphics[angle=0,scale=0.43]{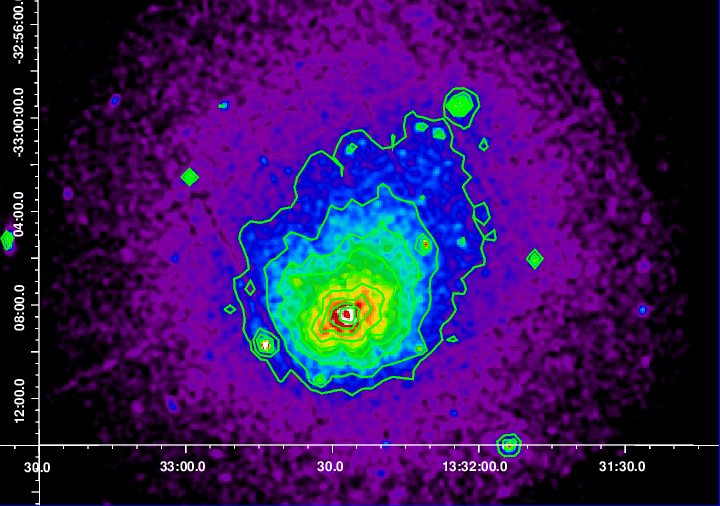}
\hspace{0.6truecm}
\includegraphics[angle=0,scale=0.44]{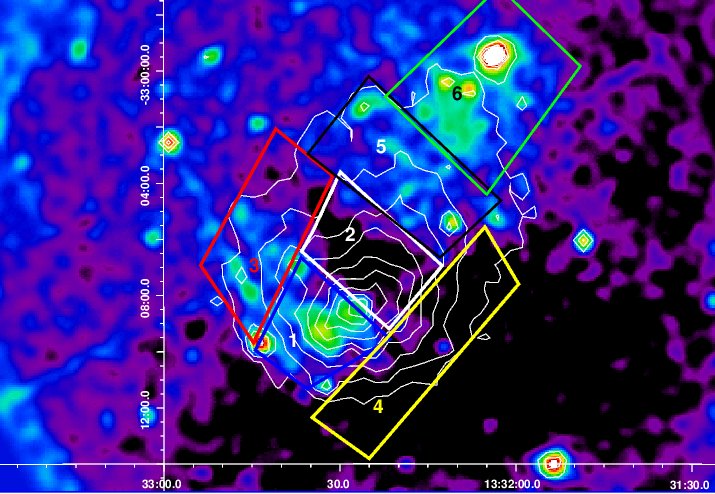}
\caption{{\it Left panel:} XMM X--ray brightness distribution of A\,3560 
(colour)
obtained using the MOS1 MOS2 and pn cameras with X--ray contours overlaid 
to help the eye. A 3 pixel smoothing is applied.
{\it Right panel:} Residual image of the X--ray surface brightness
distribution with respect to an azimuthally symmetric model in colour, 
with X--ray intensity contours (white) and regions for spectral analysis
overlaid (see Sect. 5.1 and Table 5).} 
\label{fig:fig5}
\end{figure*}
%

The result of our residual analysis, obtained with the circular symmetry 
centred on the X--ray peak, is shown in the right panel of Fig. \ref{fig:fig5}. 
The most remarkable feature in that image is the shape of the excess emission 
in the
southern part of the cluster: it starts from the X--ray surface brightness 
peak, and peaks just south--east of the cluster centre, following a 
clockwise spiral--like shape, as clearly visible in region 1.
This feature is visible in all residual images we extracted under different 
geometries, as well as in an unsharp mask analysis performed as further check, 
and therefore we regard it as reliable.
The shape of this feature recalls the spiral residuals which have 
been observed in a number of clusters 
(e.g, Lagan\'a et al. \cite{lagana08} and \cite{lagana10}), 
and in simulated residual images of sloshing clusters (e.g., 
Ascasibar \& Markevitch \cite{ascasibar06}, Roediger et al. \cite{roediger11}, 
\cite{roediger12a} and \cite{roediger12b}). 
They are interpreted as the signature of gas sloshing induced in the 
central cluster regions by minor mergers. However, the shape of this
excess is not unambiguous and its interpretation is not
straightforward: it may also be due to gas of a merging subcluster
stripped during the interaction.
Another remarkable feature of our residual image is the presence of 
diffuse excess emission at $\sim~8^{\prime}$ north--west from the cluster 
centre, suggestive of substructure (included in region 6 in the right panel 
of Fig. \ref{fig:fig7}). We will discuss the
possible interpretation of both features in Sec. \ref{sec:interpretation}.

We used the residual image to select regions for spectral analysis,
shown in the right panel of Fig. \ref{fig:fig5}.
Regions 1 and 3 were chosen to follow the spiral--like surface brightness 
excess in the inner and outer parts, and regions 2, 4 and 5 as control 
regions. Region 6 follows the north--western elongation and corresponds
to the location of the excess emission.
For each region, we extracted and analized spectra 
as described in Sec. \ref{sec:xray_obs_xmm}, the best fit parameters are 
reported in Table \ref{tab:spectra}. 
The cluster does not show significant temperature variations and we do
not have any indication of a cool core, which may support the
suggestion from the X--ray image that A3560 is not a relaxed system. 
Region 6 shows a slightly lower temperature compared to the cluster average, 
although the difference is only mildly significant.
The absence of significant temperature variations is confirmed
also by the temperature map we obtained from {\it XMM-Newton} data
using the technique in Rossetti et al. (\cite{rossetti07}). We do not
show the map here since we could not recover significant temperature
structure: best fit values in each bin are all in the range
$3.1-3.9$ keV well consistent within each other and with the values in
Table \ref{tab:spectra}, since typical error bars in the {\it XMM-Newton}
temperature map are of the order $0.3-0.4$ keV.

An {\it XMM--Newton}  temperature analysis of A\,3560 was performed 
also by Hudaverdi et al. 
(\cite{hudaverdi10}). They also found evidence for a north--western cold peak, 
with T$\sim$ 3.0 keV, which they proposed to be a sub--cluster. 
Compared to our analysis, their cold peak is more prominent with respect 
to the average cluster temperature, however they did not include error 
estimates, therefore a detailed comparison with our results is not possible. 

We performed a metal abundance analysis in the same regions shown in  
Fig. \ref{fig:fig5} . Only the central part of the ``spiral'' 
(region 1) shows a significant excess with respect to the cluster mean. It 
is interesting to note that region 2 is located close to the cluster
centre, too, but it features a lower metallicity, thus the excess
observed in region 1 cannot be explained with the metal abundance peak
typically observed in the central region of cool core clusters. A non--central 
metal abundance excess, located in a region of enhanced surface 
brightness (and therefore with low entropy) can be interpreted as a
``cool core remnant'' (Rossetti \& Molendi \cite{rossetti10}), 
i.e. what remains of the cluster cool core after a merger. Alternatively, 
the enriched gas could also belong to a merging subcluster 
(Sec. \ref{sec:interpretation}).


\begin{table}
\centering
\caption[]{Results from the X--ray spectral analysis.}
\begin{tabular}{lcc}
\hline
\hline\noalign{\smallskip}
Region &    Temperature (keV)        & Metal Abundance (solar) \\
\hline
1  (blue)  &   $ 3.48\pm0.07 $  &  $0.38 \pm 0.04$ \\
2  (white) &   $ 3.75\pm0.10 $  &  $0.21 \pm 0.03$ \\
3   (red)  &   $ 3.49\pm0.13 $  &  $0.15 \pm 0.05$ \\ 
4 (yellow) &   $ 3.38\pm0.10 $  &  $0.21 \pm 0.04$ \\            
5   (black)&   $ 3.61\pm0.10 $  &  $0.15 \pm 0.04$ \\
6   (green) &  $ 3.30\pm0.10 $  &  $0.20$ (fixed)  \\
\hline
\end{tabular}
\label{tab:spectra}
\end{table}

\begin{figure*}[htbp]
\centering
\includegraphics[angle=0,scale=0.7]{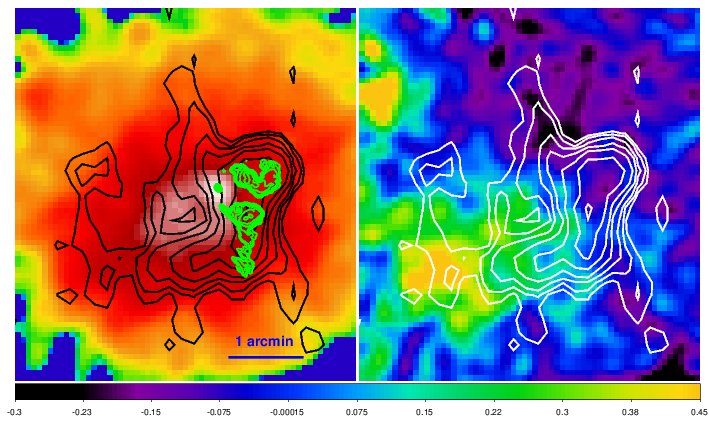}
\caption{{\it Left panel:} XMM X--ray brightness distribution of A\,3560 
(colour) 
with radio contours of J\,1332--3308 overlaid. 
Black contours are the GMRT 240 MHz image (same contours and resolution 
shown in the right panel of Fig. 2);
green contours are the VLA 8.4 GHz image (same contours and resolution shown
in Fig. 1). 
{\it Right panel:} Zoom on the XMM residual map (colour) with GMRT 240 MHz 
contours of J\,1332--3308 overlaid in white.}
\label{fig:fig6}
\end{figure*}
%

\subsection{Connection between radio and X--ray emission}

We compared the distribution of the radio emission of  J\,1332--3308 with the
characteristics of the ICM in the central region. In the left panel of
Fig. \ref{fig:fig6} we show a zoom of the XMM image (shown in Fig. 5) with
radio contours of J\,1332--3308 overlaid. The X--ray image has been smoothed
with a gaussian function with FWHM=30$^{\prime\prime}$ . 

The emission peak in the XMM image  is coincident with the radio active 
nucleus of the BCG, as clear from the 8.4 GHz radio contours overlaid.
Moreover, the X--ray/radio overlay clearly shows that the extended radio 
emission imaged at 240 MHz east of the active dumb--bell nucleus follows 
the regions where the X--ray emission is brighter. This is clearly seen 
also in the right panel of Fig. \ref{fig:fig6}, which zooms into the central
region of the residual map: the southern region of the diffuse radio 
emission and the spur follow the central part of the spiral--like 
excess discussed in Sec. \ref{sec:icm}.
At a qualitative level, the X--ray/radio comparison suggests that the 
steep spectrum components of the radio emission of J\,1332--3308, i.e. 
the diffuse component, the spur and the northern extension, are strongly 
influenced by the ICM at the cluster centre.

\begin{figure*}[htbp]
\centering
\includegraphics[angle=0,scale=0.40]{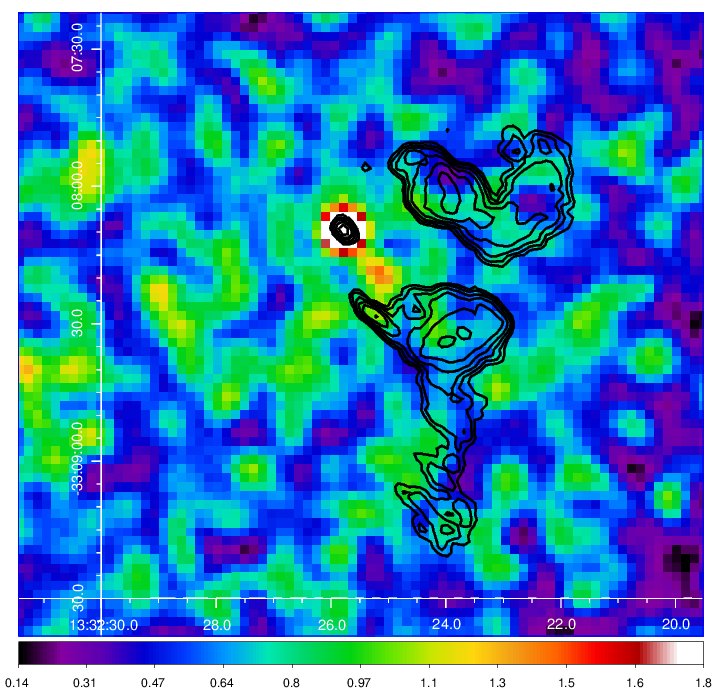}
\hspace{0.3cm}
\includegraphics[angle=0,scale=0.52]{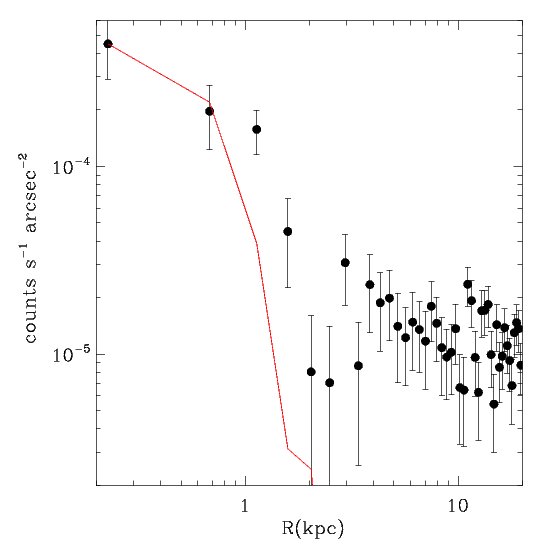}
\caption{{\it Left panel: Chandra} image of the central $2.5\times2.5^{\prime}$ 
of A3560. We binned the image with a pixel size $2^{\prime\prime}$ and it has 
been further smoothed for displaying purposes with a FWHM of 3 pixels. $8.4$ 
GHz contour are overlaid. {\it Right panel:} Surface brightness profile of 
the candidate corona compared with the PSF (red line).}
\label{fig:fig7}
\end{figure*}

We analyzed the central part of the cluster in more detail by means of 
a high resolution image (Fig.\ref{fig:fig7}, left panel) obtained from our 
more recent $Chandra$ observation: the image is very noisy because of the
short exposure time and faintness of the cluster, but we can still recognize 
a bright rim, corresponding to the more central part of the excess in the 
residual map, and an X--ray source associated with the north--eastern nucleus 
of the dumb--bell galaxy.
This X--ray source looks more extended than other point--like sources in the 
field, therefore we extracted a surface brightness profile from the
full resolution image and compared it
to the $Chandra$ PSF.\footnote{We calculated the Point Spread Function at 
the location of the source at $E=1$ keV, using the Chandra Ray Tracing 
(ChaRT) tool.} The result is given in the right panel of Fig. \ref{fig:fig7}, 
which shows that the X--ray source is qualitatively more extended than the PSF.

From the surface brightness profile, shown in Fig. 7 (right), we subtracted 
a model composed by the PSF plus a constant to account for the cluster 
emission, 
which we estimated fitting the data between 5 and 20 kpc. Below 5 kpc the 
data exceed the model at more than 6$\sigma$, therefore this source can be 
considered extended according to Sun et al. (\cite{sun07}). 
Unfortunately the low number of counts in
the $Chandra$ observation (34 in the $0.5-2$ keV energy range) does not 
allow a full spectral characterization, 
as it is consistent both with a power--law and with a thermal model. 
We applied the spectral criterion proposed by Sun et al. (\cite{sun07}) for 
faint X-ray sources: fitting the spectra with a power--law model the 
$3\sigma$ lower limit of the spectral index is larger than $2.4$, thus the 
spectrum is more similar to a thermal model with an insignificant iron L--shell
hump than to the flatter spectra of AGNs. Therefore, we classify this
source as a candidate low--luminosity X--ray corona. Assuming that all
the emission from this source is associated with the corona, we provide
an upper limit for its luminosity in the $0.5-2$ keV energy range,
$L_{max}=3.3 \times 10^{40}\rm{ergs}\, \rm{s}^{-1}$, which is
consistent with the luminosity of the faint coronae in Sun (\cite{sun09}).

%
%

\subsection{Galaxy density distribution}

Very little redshift information is available in this region of the
sky. Willmer et al. (\cite{willmer99}) derived the velocity 
for 33 galaxies and obtained controversial results with respect to 
possible subclustering.

In order to describe the dynamical state of A\,3560, we collected the
redshift data available in the literature.
Using the NED database, we found 285 redshifts, mainly from Willmer et al.
(\cite{willmer99}) and from the WING Survey (Cava et al. \cite{cava09}).
We checked the consistency of the two databases using the information on
the 26 common galaxies, and obtained a negligible sistematic shift, i.e.
19 km s$^{-1}$. The external error on the redshift determination is 
relatively small, too, i.e. 44 km s$^{-1}$. 

We computed the cluster mean velocity (v) and velocity dispersion 
($\sigma_{\rm v}$) with the bi--weight estimators (Beers et al. \cite{beers90}),
and we found v=14656 km s$^{-1}$ and $\sigma_{\rm v}=785$ km s$^{-1}$ respectively.
The value of $\sigma_{\rm v}$ is well consistent with the X--ray temperature 
derived from the {\it XMM--Newton} observations (Table 5). 
Statistical tests do not show significant deviation from the gaussianity.
However, our redshift sample is biased against the central part of the 
cluster, and does not allow a proper analysis in the region we are dealing
with here.

\begin{figure}[htbp]
\centering
\includegraphics[angle=0,scale=0.47]{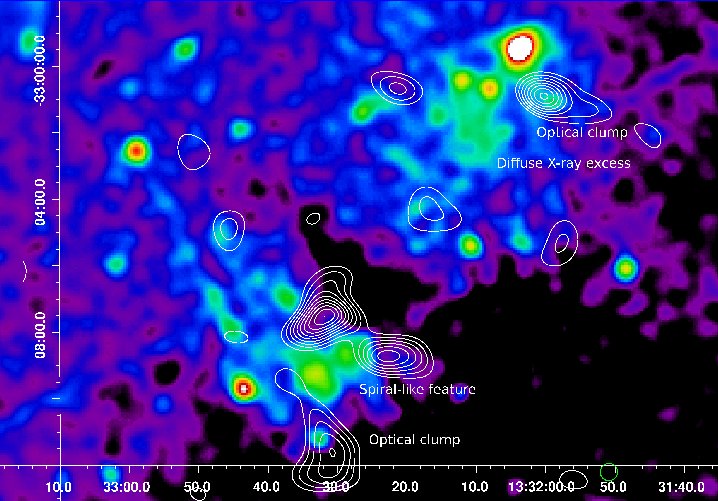}
\caption{Distribution of the optical galaxies as derived from the APM
catalogue (white contours) overlaid on the {\it XMM--Newton} surface 
brightness residual image, shown in colour (same image as right panel
of Fig. \ref{fig:fig7}). The labels show the location of the two clumps 
referred to in Sect. 5.3 and of the X--ray brightness excess features 
(Sect. 5.1).}
\label{fig:fig8}
\end{figure}

In the X--ray study, we identified a possible substructure north--west
of the core which could be undergoing a merger with the main cluster
in the plane of the sky, very difficult to detect in the 
redshift space. We then performed a bi--dimensional analysis in order 
to detect possible sub--clumps. 
The WING Survey (Valentinuzzi et al. \cite{valentinuzzi09}) 
is in principle the best suited database, but unfortunately an instrumental 
rib is located just on the dominant galaxy, preventing a substructure
analysis in this region. We thus used the central part of the APM survey 
(Irvin \cite{irvin85}), which is limited to R=21.5. 
To detect substructure, we used the Adaptive Kernel method (DEDICA algorithm) 
as implemented by Pisani (\cite{pisani96}),  and applied to clusters by 
Bardelli et al. (\cite{bardelli98b}). 
Our results are given in Fig. \ref{fig:fig8} which shows the substructure 
analysis over whole cluster field as obtained from the APM catalogue 
(white contours) overlaid on the X--ray surface brightness redisuals.

The distribution of the galaxies is bimodal at the cluster centre, with
two peaks of comparable mass. The radio BCG is located just between 
the two condensations. Two further significant clumps (labelled in Fig. 
\ref{fig:fig8}) have been detected. 
One is located close to the diffuse X--ray brightness residual 
emission, while the other is close to the spiral--like X--ray surface 
brightness excess at the cluster centre (see Fig. \ref{fig:fig8}, and 
regions 6 and 1 respectively in the right panel of Fig. \ref{fig:fig5}).
None of these two optical clumps is coincident with the X--ray
features. However, misplacements between gas and galaxies
have already been observed in some cluster mergers, as expected due to 
the different behaviour of the collisional (gas) and the non--collisional 
(galaxies and dark matter) components.
\\
Although not used in the analysis of the cluster centre, the WINGS 
bidimensional distribution of galaxies supports the existence of these two 
clumps, and confirms that they are both off--placed with respect to the 
features detected in the X--ray band.

%

\section{Discussion}

In this paper we present a detailed radio, X--ray and optical analysis  
of the cluster A\,3560 and of the BCG at its centre, associated with the
radio galaxy J\,1332--3308.
The cluster is located at the southern periphery of the A\,3558 cluster 
complex, in the core of the Shapley Concentration. 
Our study is based on proprietary high sensitivity radio observations -- 
GMRT at 1280, 610 and 240 MHz; VLA at 1.4, 4.86 and 8.46 GHz; ATCA at 2.3 
GHz-- and on proprietary {\it XMM--Newton} and {\it Chandra} data, while the 
optical analysis was performed using literature and archival information.

The multifrequency radio images show that the radio emission associated with 
the brightest cluster galaxy at the centre of A\,3560 is enigmatic, and cannot 
be simply interpreted as an active radio galaxy providing the whole 
relativistic plasma. Moreover, our optical and X--ray analysis suggest that 
the cluster has recently undergone a minor merger.
Our results can be briefly summarized as follows:

\begin{itemize}

\item{} The radio emission associated with the cluster BCG is characterized
by an AGN component, with an active core and two lobes, and by
extended diffuse emission. The latter is best imaged at frequencies below 
1 GHz, it is asymmetric with respect to the active radio galaxy, and extends 
eastwards with respect to the location of the BCG. The overall size of the 
diffuse radio emission is $\sim$220 kpc. 
A detailed radio spectral study shows that the diffuse emission has a very 
steep spectrum, with $\alpha>2$ for $\nu \ge 610$ MHz. Its radio plasma is 
aged (t$_{\rm diffuse} \sim 100$ t$_{\rm lobes}$) and is not currently fed by 
the active nucleus. 

\item{} A morphological comparison between the emission of the
radio galaxy and the X--ray surface brightness of the ICM as imaged by
{\it XMM--Newton} shows a clear correlation of features, the radio
emission following the X--ray excess emission.

\item{} The radio core is associated with the northern nucleus of the BCG
and with an X--ray extended source in the {\it Chandra} image, which we 
classify as a low--luminosity corona. The radio and X--ray properties of the 
BCG in A\,3560 are consistent with those of coronae in 
poor cluster BCGs, as found in other studies (Sun \cite{sun09} and references 
therein).

\item{} The temperature of the ICM, as derived from {\it XMM--Newton} 
observations, is kT$\sim$3.5 keV. The overall shape of the X--ray
surface brightness distribution is elongated in the south--east/north--west
direction, with a positional shift between the centre of the inner and 
outer isophotes, suggestive of an unrelaxed dynamical state.
A residual image shows the presence of a spiral structure at the cluster 
centre, which appears to be associated with the diffuse 
240 MHz radio emission.
The analysis of the gas temperature is consistent with the presence of a 
subcluster, located $\sim 8^{\prime}$ north--west of the cluster centre.
Finally, the ICM in the central regions featuring a surface brightness 
excess is more enriched in metals than the other central parts of the 
cluster. This may suggest that A\,3560 has undergone a cool core phase in an
earlier stage of its evolution (Rossetti \& Molendi \cite{rossetti10}).

\item{} A bidimensional optical analysis shows that the galaxies in the 
central cluster region have a bimodal distribution, the two groups having 
comparable density. Two more clumps are identified: one 
is located $\sim 8^{\prime}$ north--west of 
the cluster centre, and it is the possible optical counterpart of the gas 
feature 
detected from the X--ray analysis, the other is located close to the
surface brightness excess south of the cluster centre.

\end{itemize}

\subsection{Core sloshing at the centre of A\,3560 and its consequences on 
the radio emission}
\label{sec:interpretation}

The interpretation of the dynamical state of A\,3560 from the 
X--ray and optical data (Sec. \ref{sec:icm}) is not
straightforward. Our spectral and imaging analysis suggest an unrelaxed 
dynamical state but the lack of significant thermodynamic features, due 
to the isothermality of the cluster, does not help in recovering the merging 
history of this cluster. The only significant variations we detected in the 
ICM are in the residual image, which shows substructure, possibly associated 
to a galaxy concentration, to the north--west, and a spiral--like excess, 
enriched in metals, close to the centre of the cluster. 

We propose a scenario 
where A\,3560 is undergoing an off--axis minor merger event, which may have 
induced gas motions in the cluster core, leaving a signature in the form of a
spiral residual. This scenario is similar to the sloshing phenomenon 
(Ascasibar \& Markevitch \cite{ascasibar06}) which usually induces cold fronts 
in the ICM of the main cluster. We do not observe cold fronts in 
A\,3560: we searched for arc--like features and discontinuities in the 
{\it XMM-Newton} image and profiles, but we could not find any significant 
indication. However, we cannot rule out the presence of cold fronts in the 
inner part of the cluster ($\sim 1^{\prime}$): unfortunatley, our 
{\it XMM-Newton} data are not well suited to resolve the central region, 
and the superior angular resolution of {\it Chandra} is hampered by the poor 
statistics of the observation. 
Moreover, A\,3560 is a non cool core cluster, different from the 
clusters with steep entropy profile which usually host sloshing cold fronts 
(e.g. Ghizzardi et al. \cite{ghizzardi10}), although with a few notable 
exceptions (e.g. A2142, Markevitch et al. \cite{markevitch00}, 
Rossetti et al. \cite{rossetti13}). 
Simulations of off-axis mergers in non cool core clusters (i.e. clusters 
without a steep entropy profile) show that these mergers induce complex 
motions in cluster cores and asymmetric gas structure that would be 
seen in the residual maps but do not generate cold fronts (see Fig.\,12 of 
Ascasibar \& Markevitch \cite{ascasibar06}). We suggest a similar 
``sloshing-like'' scenario for A\,3560.
The identification of the perturber is not trivial, and projection
effects may further complicate the interpretation of the observations.
Our X--ray and optical 
analysis suggest that the group located $\sim 8^{\prime}$ north--west of
the cluster core, in the direction of A3558, is the most likely
candidate. 

Alternatively, the excess close to the cluster centre, enriched in 
metals, may be associated with stripped gas from a  colliding subcluster.
A second optical condensation is present just south of the cluster 
core, close to the feature visible in the X--ray surface 
brightness residual image 
(see Fig. 8 and Sect. 5.3). It is thus possible that A\,3560 is a multiple 
merger, and that the X--ray features close to the centre are a direct
effect of the interaction, rather than an indirect one as in the
sloshing--like scenario. 

The properties of the radio emission associated with the BCG support and
strengthen the idea of dynamical activity at the cluster centre.
The diffuse emission is aged, and not currently fed by nuclear activity. 
Our estimates of the radiative ages (Sect. 4.2) are consistent with the
idea that this older plasma may have been deposited in the ICM by a 
previous cycle of activity of the BCG, and either left behind by the motion
of the BCG around the cluster centre, or affected by the interaction with
the ICM environment. In the former scenario, and assuming that the
ICM is at rest in the potential well of the cluster, the ring--shaped 
morphology of the spur (right panel of Fig. 2) contains information on the 
trajectory of the BCG, in its orbit around the cluster centre. The spectral 
steepening of the spur going from west to east (right panel of Fig. 4) would 
suggest an anticlock--wise orbital motion. The size and shape of the possible 
trajectory (in the plane of the sky) is $\sim 4.3^{\prime}$, and assuming a 
value for the radiative age of the diffuse emission of the order of 
t$_{\rm rad}\sim2\times10^8$ yr (Sect. 4.2) , we obtain an indicative value
for the velocity of the BCG of $\sim$ 1000 km s$^{-1}$, which is very high 
for cluster BCGs. The most extreme values reported
so far for this class of objects are of the order of 6--700 km s$^{-1}$ 
(e.g., A\,3653, Pimbblet et al. \cite{pimbblet06} and references therein).
Our estimate should be taken with care, considering the uncertainties 
related to the age estimate of the diffuse radio emission, and the 
over--simplified assumptions on the galaxy velocity and trajectory 
(see below).

An alternative possibility for the origin of the radio emission 
of J\,1332--3308 is that the aged diffuse emission is dragged and wrapped 
up by the the gas motion at the cluster centre, as the radio/X--ray 
overlays in Fig. 6 strongly suggest. This association is naturally 
explained in the sloshing--like scenario, while it would be only a chance 
superposition if the gas excess was due to a merging substructure, unrelated 
with the radio-emitting BCG.  
For this reason, we favour the former scenario for the merging history of
this cluster.

If the north--western group is indeed the perturber of A\,3560, then the 
shape of the diffuse radio emission, and its extension east of the location 
of the BCG, suggest that the minor merger might have taken place from 
south--east to north--west. 
On the other hand, the orientation of the lobes (Fig. 1) shows that at 
the location of the active nucleus the radio plasma is experiencing a motion 
in the east--west direction: it is impossible to discriminate between an 
eastward motion of the BCG, and a velocity field westwards in the ICM. 
Nevertheless, the overall morphology of the radio emission indicates the 
presence of a complex velocity field in the inner cluster regions. 
Ascasibar \& Markevitch (\cite{ascasibar06}) proposed that both gas sloshing
and the peculiar velocities of cD galaxies can be generated by the same
minor merger events. It is possible that the bent morphology of the jets
and lobes of J\,1332--3308 is the signature of the oscillations of the
BCG in A\,3560 as consequence of the minor merger. As a final remark,
we point out that according to Ascasibar \& Markevitch (\cite{ascasibar06}) 
the trajectory of the dark matter and gas peak in the plane of the merger 
tend to form a circle (Fig. 4 in their paper), which is consistent 
with the trajectory derived for the BCG on the basis of the diffuse
emission.

Most likely both an orbital motion of the BCG around the cluster centre, 
and gas sloshing in the ICM are responsible for the peculiar radio emission
at the centre of A\,3560. Therefore, we can safely conclude that the
overall morphology and properties of J\,1332--3308 are strongly influenced
by the interaction with the external medium, and are the result of the
dynamical activity at the cluster centre.

\subsection{The low--luminosity X--ray corona of the BCG}

Another important piece of information is the presence of a small ($<$ 5 kpc) 
low--luminosity corona in the  inner region of the BCG. 

Coronae are common in BCGs both in rich and poor environments, and can be
considered scaled versions of cluster cool cores (Sun \cite{sun09}). The 
1.4 GHz radio luminosity of J\,1332--3308, as well as its morphology, are 
consistent with the known high radio luminosity corona galaxies. Moreover, 
the radio power of the BCG and the X--ray properties of A\,3560 are consistent 
with the corona class objects studied in Sun (\cite{sun09}). 
Indeed if we overlay the upper limit for the X--ray luminosity of the corona  
and the radio luminosity of J\,1332--3308 in Fig. 1 of Sun (\cite{sun09}), 
this source is not an outlier in the distribution and falls well within the 
region of low X--ray luminosity high radio power objects in the corona class.
Finally, the radio emission of J\,1332--3308 shows a clear gap between the 
nucleus and the lobes 
(Fig. 1 and Section 3.1), as it has been found in other radio loud galaxies 
with a corona, both in rich clusters (e.g., NGC\,4874 in Coma, Sun et al. 
\cite{sun07}) and in groups (e.g., AWM\,04, O'Sullivan et al. 
\cite{osullivan11}). All the above provides further support to the suggestion
that indeed the BCG in A\,3560 hosts a low--luminosity corona.

Sun et al. (\cite{sun07}) showed that the central parts (i.e. $\ltsim$ 4 kpc) 
of coronae may survive gas stripping in unrelaxed dense environments, such 
as minor merger induced sloshing motions, and even major mergers. The long 
time scale of gas sloshing (of the order of the Gyr), combined with the cool 
temperatures of coronae, suggest that their lifetime is long, and that 
some mechanism, beyond the possible replenishment from stellar mass loss,
must be present to preserve them. Our data are not suitable to 
address this point, however we can safely conclude that the detection
of a corona in the cD galaxy at the centre of A\,3560 is another piece of 
evidence in favour of the presence of mechanisms preventing the evaporation 
of galactic coronae.

\section{Conclusions}

The radio galaxy J\,1332--3308, associated with the north--western nucleus
of the dominant cluster galaxy in A\,3560, is very peculiar. It consists 
of an active component in the form of a core, jets and radio lobes, and of 
aged diffuse emission extending over a scale of $\sim$220 kpc.
\\
The multiband X--ray and optical analysis, performed to address the origin 
of this radio galaxy, supports the idea that A\,3560 is a very interesting 
example of radio AGN/ICM interaction in a minor merger.

The X--ray observations suggest a minor merger scenario between 
the main cluster and a group:
the surface brightness X--ray residual image shows the presence of a
spiral--like feature in the central region, a typical signature of gas 
sloshing. Moreover, excess emission is detected $\sim 8^{\prime}$ north--west
of the cluster centre, suggestive of a sub--cluster, whose presence is 
supported also by the temperature analysis. 

The galaxy distribution is bimodal at the cluster centre. A third 
condensation is found north--west, in the direction of the Shapley 
Concentration core, at the location of the X--ray sub--cluster. Finally,
another optical clump is located south of the cluster centre, close
to the spiral--like structure seen in the residual X--ray image.

A clear connection is found between the spiral--like feature detected in the
residual X--ray image and the diffuse, aged (t$\sim 1.5\times10^8$ yr) 
component of the radio emission, whose morphology is most likely affected 
by the gas motion in the cluster central region.
The BCG is currently active in the radio band, with an FRI radio galaxy  
associated with the north--western 
nucleus. The radio lobes have a radiative age of the order of 1--2 Myr.
Their orientation has no connection with the older diffuse emission, and  
possibly reflects the relative motion between the BCG and the local ICM.
A low--luminosity corona was detected in the inner region (\ltsim ~ 4 kpc)
of the radio loud nucleus of the BCG, but unfortunately, our {\it Chandra} 
exposure is too short to detect a possible extension, which would help to 
derive information on the current motion of the BCG with respect to the ICM.

\medskip

{\it Acknowledgements.}
The authors thank the anonymous referee, whose comments and suggestions
improved the paper. 
We thank M. Murgia for careful reading of the manuscript and fruitful 
discussion. M.R. acknowledges support by the program 
``Dote Ricercatori per lo sviluppo del capitale umano nel Sistema
Universitario Lombardo'', FSE 2007-2013, Regione Lombardia.
S.G. acknowledges the support of NASA through Einstein Postdoctoral
Fellowship PF0--110071 awarded by the Chandra X--ray Center (CXC), which
is operated by the Smithsonian Astrophysical Observatory (SAO).
We thank the staff of GMRT for their help during the observations. GMRT is
run by the National Centre for Radio Astrophysics of the Tata Institute of
Fundamental Research. 
The National Radio Astronomy Observatory (NRAO) is a
facility of the National Science Foundation operated under cooperative
agreement by Associated Universities, Inc. The Australia Telescope 
Compact Array is part of the Australia Telescope National Facility which is
funded by the Commonwealth of Australia for operation as a National
Facility managed by CSIRO.

\end{document}